\documentclass[11pt,twoside]{article}

\usepackage{asp2006_galagb}
\usepackage{epsf}

\usepackage{graphicx}  % SH

\begin{document}
\title{Headwind: Modelling Mass Loss of AGB Stars, \\ Against All Odds}   %%% Fill in title
\author{Susanne H{\"o}fner}   %%% Fill in author names
\affil{Dept. of Astronomy \& Space Physics, Uppsala University,
Box 515, SE-75120 Uppsala, Sweden}    %%% Fill in author affiliations

\begin{abstract} %%% Abstract to run on from here.
The intricate interplay of atmospheric shock waves and a complex,
variable radiation field with non-equilibrium dust formation
presents a considerable challenge to self-consistent modelling of
atmospheres and winds of AGB stars. Nevertheless it is clear that
realistic models predicting mass loss rates and synthetic spectra
are crucial for our understanding of this important phase of stellar
evolution. While a number of questions are still open, significant
progress has been achieved in recent years. In particular,
self-consistent models for atmospheres and winds of C-stars have
reached a level of sophistication which allows direct quantitative
comparison with observations. In the case of stars with C/O $< 1$,
however, recent work points to serious problems with the dust-driven
wind scenario. This contribution analyzes the basic ingredients of
this scenario with analytical estimates, focusing on dust formation,
non-grey effects, and differences between C-rich and O-rich
environments, as well as discussing the status of detailed dynamical
wind models and current trends in this field.
\end{abstract}

\section{Introduction}

Mass loss is one of the most pronounced features of AGB stars,
influencing both their evolution and their observable properties in
a decisive way. The most commonly accepted scenario is that of a
pulsation-enhanced dust-driven wind: stellar pulsation causes
atmospheric shock waves which intermittently lift gas above the
stellar surface, creating dense, cool layers where solid particles
form. The dust is accelerated away from the star by radiation
pressure, dragging the gas along. The composition of the grains is
determined by the relative elemental abundances in the atmosphere,
with C-rich stars forming mostly amorphous carbon grains, and O-rich
objects producing silicate dust.

Current knowledge suggests that winds of C-rich AGB stars are rather
well understood. For their numerous O-rich counterparts, however,
recent work demonstrates that the opacities of silicate grains are
insufficient for driving winds (Woitke 2006a and this volume),
contrary to previous expectations. This problem becomes apparent
with the introduction of frequency-dependent radiative transfer in
time-dependent models with a detailed description of dust formation.
Ironically, this important step towards more realistic models which
brought C-rich models into good agreement with observations, causes
serious problems with the wind mechanism in O-rich stars.

The core of this issue can be understood with comparatively simple
analytical arguments. A significant part of this review is devoted
to isolating the crucial ingredients and analyzing them with
simplified analytical descriptions, in particular non-equilibrium
dust formation, with special attention to the influence of non-grey
effects and differences between C-rich and O-rich stars. These
simple estimates are compared to the results of existing numerical
models, rounding up with an overview of the present status of such
detailed models, and current trends.

Detailed comparisons of the input physics of various numerical
models and summaries of the historical developments in this field
can be found in earlier reviews, e.g., by Willson (2000), Woitke
(2003) and H{\"ofner} (2005).

\section{Basic ingredients and characteristic numbers}

Dust grains forming in the extended atmospheres of AGB stars
condense under non-equilibrium conditions with temperature acting as
a threshold, prevailing densities and abundances determining the
efficiency of grain growth, and pulsation and shock waves setting
the time scales. In other words, condensation will usually be
incomplete, and the dust-to-gas ratio is not a simple function of
abundances.

\subsection{Grain temperature and condensation radius}

A prerequisite for condensation or survival of existing grains are
temperatures below the stability limit of a specific condensate. In
principle, a distinction between the respective temperatures of
ambient gas and grains has to be made. In practice, however, gas
temperatures tend to be lower than grain temperatures at a given
point, and therefore the limit will usually be defined by the grain
temperatures which, in turn, are given by radiative equilibrium,
i.e.
\begin{equation}%\label{}
    \int_{0}^{\infty} \kappa_{\lambda} B_{\lambda}(T_d) \, d \lambda
      =  \int_{0}^{\infty} \kappa_{\lambda} J_{\lambda} \, d \lambda
\end{equation}
Using the assumption of a Planckian radiation field (with stellar
surface temperature $T_{\ast}$ at $R_{\ast}$), geometrically diluted
according to distance from the stellar surface
\begin{equation}%\label{}
    J_{\lambda} = W(r) \, B_{\lambda}(T_{\ast})  \qquad
    W(r) = \frac{1}{2} \left( 1 - \sqrt{ 1 - \left( \frac{R_{\ast}}{r} \right) ^2 } \right)
\end{equation}
in the equation of radiative equilibrium, the grain temperature
$T_d$ as a function of distance from the star can be estimated if
the wavelength-dependent opacity of the dust particles
$\kappa_{\lambda}$ is known. The latter can often be approximated in
the critical wavelength region around the stellar flux maximum by
$\kappa_{\lambda} \propto \lambda^{-p}$ where the exponent $p$ is
characteristic of the grain material, leading to
\begin{equation}%\label{}
    W(r) = \frac{\int_{0}^{\infty} \kappa_{\lambda} B_{\lambda}(T_d) \, d \lambda}
             {\int_{0}^{\infty} \kappa_{\lambda} B_{\lambda}(T_{\ast}) \, d \lambda}
         = \left( \frac{T_d}{T_{\ast}} \right)^4
             \frac{\kappa_{\lambda}^{\rm Planck}(T_d)}{\kappa_{\lambda}^{\rm Planck}(T_{\ast})}
         = \left( \frac{T_d}{T_{\ast}} \right)^{4+p}
\end{equation}
where the superscript 'Planck' denotes a Planck mean of the opacity
(see, e.g., Lamers \& Cassinelli 1999 for details). The condensation
radius $r_c$, defined as the point where the grain temperature $T_d$
is equal to the condensation temperature $T_c$ (stability limit) of
the respective grain material, is therefore given by
\begin{equation}%\label{}
    \frac{r_c}{R_{\ast}} = \frac{1}{2} \left( \frac{T_c}{T_{\ast}} \right)^{- \frac{4+p}{2}}
\end{equation}
where we have used the approximation $W(r) \approx (R_{\ast}/2r)^2$
for $r \gg R_{\ast}$.

Therefore, to determine the location where condensation may start
for a particular material, we need to know its condensation
temperature and the dependence of the opacity on wavelength around
the flux maximum. For amorphous carbon grains we have $T_c \approx
1500\,$K and $p \approx 1$. Assuming a stellar surface temperature
of $T_{\ast} \approx 3000\,$K we obtain $r_c/R_{\ast} \approx 3$
which is in good agreement with detailed frequency-dependent models
where dust usually forms at $2-3\,R_{\ast}$.

\begin{figure}[t]
\vspace{-1.0cm} \hspace{0.75cm}
\includegraphics[width=12cm, angle=270]
{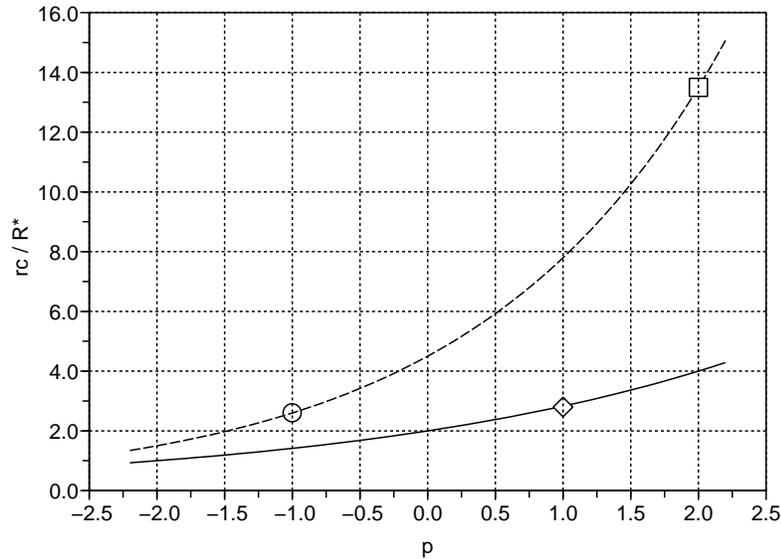} \vspace*{-3.5 cm} \caption{Condensation radius (in
units of stellar radius) as a function of $p$. The lower curve
corresponds to $T_c/T_{\ast} = 1/2$ (representative of carbon
grains; full line) and the upper curve to $T_c/T_{\ast} = 1/3$
(characteristic of silicates; dashed). The symbols show the
approximate location of various materials in this diagram (diamond:
amorphous carbon; circle: Mg$_2$SiO$_4$; square: MgFeSiO$_4$).
Models with grey radiative transfer (or grey dust opacities)
correspond to $p=0$.} \label{f_rcond}
\end{figure}

For silicate grains the picture is more complicated. To illustrate
the problem, we consider olivine, i.e. a material with the
composition Mg$_{2x}$Fe$_{2(1-x)}$SiO$_4$ ($0 \leq x \leq 1$) and a
condensation temperature of $T_c \approx 1000\,$K in the relevant
density range. On the one hand, the lower value of $T_d/T_{\ast}
\approx 1/3$ (compared to $\approx 1/2$ for carbon grains) leads to
a stronger dependence of $r_c/R_{\ast}$ on $p$ (see
Fig.~\ref{f_rcond}). On the other hand, the value of $p$ varies
strongly with the relative Fe-content ($x$) of the material, with
larger (more positive) values of $p$ for iron-rich grains. For
forsterite (MgSiO$_4$), representing one end of the chemical
spectrum, lab data indicates $p \approx -1$ in the relevant
wavelength rage (see Fig.~2 in Andersen, this volume). For
MgFeSiO$_4$ (with equal amounts of Mg and Fe, corresponding roughly
to solar composition) the value is closer to $p \approx 2$.
Therefore forsterite may condense at a distance of $ \approx 3 \,
R_{\ast}$ while MgFeSiO$_4$ will probably not form closer to the
star than $ \approx 14 \, R_{\ast}$. Already these first estimates
lead us to expect that silicate grains forming in AGB stars will
tend to be iron-poor, a conclusion which is supported by other
arguments and detailed modelling (see below).\footnote{Similar
arguments hold for pyroxenes, another possibly abundant grain
material, with a composition Mg$_{x}$Fe$_{(1-x)}$SiO$_3$ ($0 \leq x
\leq 1$).}

\subsection{Dust opacity, shock waves and wind regimes}

Assuming for the moment that a certain dust species will actually
form beyond its condensation radius, will the grains significantly
contribute to driving a wind? This is, of course, a matter of the
corresponding opacity produced by the grains. If we can assume that
gravity and radiation pressure are the only relevant forces beyond
$r_c$, the (co-moving) equation of motion for a matter element can
be written as
\begin{equation}%\label{}
    \frac{du}{dt} =
       - g_0 \left( \frac{r_0}{r} \right)^2 \left( 1 - \Gamma \right)
    \qquad {\rm with} \qquad
    \Gamma = \frac{\kappa_H L_{\ast}}{4 \pi c G M_{\ast}}
\end{equation}
where $r$ is the distance from the stellar center, $u$ the velocity,
$g_0 = G M_{\ast} / r_0$ denotes the gravitational acceleration at
the (arbitrary) point $r_0$, $\kappa_H$ the flux mean opacity, and
$M_{\ast}$ and $L_{\ast}$ are the stellar mass and luminosity,
respectively (all constants have their usual meaning). The material
properties (and relative abundances) of the dust species determine
the value of $\Gamma$ which, in turn, determines the further
dynamics of the matter element. Several special cases are
immediately apparent:
\begin{itemize}
\item
If no (or very little) dust is forming, $\Gamma$ will be (close to)
zero. The matter element will follow a ballistic trajectory,
according to its initial velocity,\footnote{In the discussion here
we always assume that the velocity before reaching the dust
condensation radius is below the escape velocity. Otherwise it would
be misleading to talk about a dust-driven wind.} reaching a maximum
distance $r_{\rm max}$ of
\begin{equation} %\label{}
    \frac{r_0}{r_{\rm max}} = 1 - \left( \frac{u_0}{u_{\rm esc}} \right)^2
    \qquad {\rm where} \qquad
    u_{\rm esc} = \sqrt{\frac{2 G M_{\ast}}{r_0}}
\end{equation}
and then falling back towards the stellar surface.
\item
If $\Gamma = 1$, the r.h.s of the equation of motion vanishes. The
matter element continues to move at a constant velocity.
\item
If $\Gamma > 1$, $du/dt$ is positive, the matter is accelerated away
from the star.
\item
If $0 < \Gamma < 1$, the final fate of the matter element (escape or
fall-back) depends on the velocity at which it is moving when it
reaches $r_c$.
\end{itemize}

The dynamical behavior in the regime $0 < \Gamma < 1$ can be
investigated with the following simplified model: if we assume that
$\Gamma$ is constant outside $r_c$ (and zero inside), the factor
$\left( 1 - \Gamma \right)$ is a constant that can be multiplied
with $g_0$, resulting in an equation of motion
\begin{equation}%\label{}
    \frac{du}{dt} =
       - g_{\Gamma} \left( \frac{r_0}{r} \right)^2
    \qquad {\rm with} \qquad
    g_{\Gamma} = g_0 \left( 1 - \Gamma \right)
\end{equation}
which formally looks like the ballistic equation of motion for
$\Gamma = 0$, but with a re-scaled gravitational acceleration
$g_{\Gamma}$, and therefore a re-scaled (lower) escape velocity
$u_{\rm esc}^{\Gamma} = \sqrt{2 G M_{\ast} (1 - \Gamma) / r}$.
Consequently, the actual velocity of the matter element when
reaching $r_c$ may be above $u_{\rm esc}^{\Gamma}$ (while still
being below the escape velocity without dust), resulting in a wind.

To find the critical case that divides fall-back from outflow, we
assume that the velocity prior to condensation is only due to the
pulsation-induced shocks propagating through the atmosphere. We use
the simplified picture that a shock passing through the matter
element under consideration can be treated as an instantaneous
acceleration to a velocity $u_0$ at the point $r_0$ (which will in
practice be close to $R_{\ast}$), followed by a ballistic movement
of the shocked gas. We consider the solution of the ballistic
equation of motion ($\Gamma=0$) for $r \leq r_c$, requiring that the
velocity at the point $r_c$ matches $u_{\rm esc}^{\Gamma}$, i.e.,
\begin{equation}%\label{}
    u_0^2 - u^2(r_c) = - 2 G M_{\ast} \left( \frac{1}{r_c} - \frac{1}{r_0} \right)
    \qquad {\rm and} \qquad
    u^2(r_c) = \frac{2 G M_{\ast}}{r_c} \left( 1 - \Gamma \right)
    \,.
\end{equation}
For $0 < \Gamma < 1$ we can therefore distinguish two cases:
\begin{itemize}
  \item a pulsation-supported dust-driven wind regime for \\
  \begin{equation}
    1 > \Gamma > \frac{r_c}{r_0} \left( 1 - \left( \frac{u_0}{u_{\rm esc}} \right)^2 \right)
    \qquad {\rm or} \qquad
    1 > \frac{u_0}{u_{\rm esc}} > \sqrt{ 1 - \Gamma \left( \frac{r_0}{r_c} \right) }
  \end{equation}
  \item a 'parachute regime' (decelerated fall-back) for \\
  \begin{equation}
    \frac{r_c}{r_0} \left( 1 - \left( \frac{u_0}{u_{\rm esc}} \right)^2 \right) > \Gamma > 0
    \qquad {\rm or} \qquad
    \sqrt{ 1 - \Gamma \left( \frac{r_0}{r_c} \right) } > \frac{u_0}{u_{\rm esc}} > 0
  \end{equation}
\end{itemize}
Assuming that $r_0 \approx R_{\ast}$ and, consequently, $u_0/u_{\rm
esc} \approx \left( u_{\rm shock} / u_{\rm esc} \right)_{\rm
r=R_{\ast}}$ the condition for a pulsation-supported dust-driven
outflow can be reformulated in the more convenient form
\begin{equation}
    \left( \frac{u_{\rm shock}}{u_{\rm esc}} \right)_{\rm r=R_{\ast}} \,
      > \, \sqrt{ 1 - \Gamma \left( \frac{R_{\ast}}{r_c} \right) }
    \qquad ( 0 < \Gamma < 1 )
\end{equation}
where $u_{\rm shock}$ denotes the velocity of the shocked gas (not
to be confused with the shock amplitude).

In order to determine the value of $\Gamma$ for various grain
materials, we first need to calculate the corresponding (flux mean)
opacities. For simplicity, we assume that $\kappa_H$ can be
approximated with the small particle limit opacity of spherical
grains at $\lambda_{\rm max} \approx 1 \mu$m (flux maximum of the
star). Therefore, we have
\begin{equation}%\label{}
      \kappa_H \approx
        \frac{1}{\rho} \int_0^{\infty} a^2 \pi \, Q_{\rm ext}(\lambda_{\rm max}) \,\, n(a) \, da
        = \frac{\pi}{\rho} \, Q'_{\rm ext}(\lambda_{\rm max}) \int_0^{\infty} a^3 \, n(a) \, da
\end{equation}
where $a$ denotes the grain radius, $n(a)$ the number density of
grains and $Q_{\rm ext}$ is the extinction efficiency. For grains
small compared to the wavelength, the quantity $Q'_{\rm ext} =
Q_{\rm ext}/a$ becomes independent of the grain radius and can
therefore be taken out of the integral. The last integral represents
the fraction of a given volume which is occupied by the grains,
apart from a factor $4 \pi / 3$, and it can be rewritten in terms of
the space occupied by a monomer (basic building block) in the
condensed material times the number of monomers found in a certain
volume,
\begin{equation}%\label{}
      \int_0^{\infty} a^3 \, n(a) \, da
         = \frac{3}{4 \pi} \, V_{\rm mon} \, n_{\rm mon}
         = \frac{3}{4 \pi } \, \frac{A_{\rm mon} m_p}{\rho_{\rm grain}} \,\,
            f_c \, \varepsilon_c \, n_{\rm H}
\end{equation}
where we have expressed the monomer volume $V_{\rm mon}$ in terms of
the atomic weight of the monomer $A_{\rm mon}$ and the density of
the grain material $\rho_{\rm grain}$, and the number of monomers in
a volume $n_{\rm mon}$ by the abundance of the key element of the
condensate $\varepsilon_c$, the degree of condensation of this key
element $f_c$, and the total number density of H atoms $n_{\rm H}$
($m_p$ = proton mass). Using $n_{\rm H} = \rho / (1 + 4 \,
\varepsilon_{\rm He}) \, m_p$, we finally obtain
\begin{equation}%\label{}
      \kappa_H \approx
         0.5 \,\, \frac{A_{\rm mon}}{\rho_{\rm grain}} \,\, Q'_{\rm ext}(\lambda_{\rm max}) \,
          \,\, f_c \, \varepsilon_c \,.
\end{equation}
Below, we list the properties of different grain materials and the
corresponding $\Gamma$, assuming $L/M = 5000 \, L_{\odot}/M_{\odot}$
and a degree of condensation of the key element $f_c = 1$ (note that
the value of $\Gamma$ scales linearly with each of these two
factors). The abundance of the key element $\varepsilon_c$
corresponds to either all C not bound in CO for carbon grains (C/O =
1.5), or to Si for silicates. The data for $Q'_{\rm ext}$ is taken
from Rouleau \& Martin (1991, amorphous carbon) and J{\"a}ger et al.
(2003, silicates; cf. Figs.~1 and 2 in Andersen, this volume).

\begin{tabular}{llllll}
  & \\
  \hline
  material & $A_{\rm mon}$ & $\rho_{\rm grain}$ & $Q'_{\rm ext}$ & $\varepsilon_c$ & $\Gamma^{f_c = 1}_{L/M=5000}$ \\
   &  & [g/cm$^3$] & [1/cm] &  &  \\
  \hline
  & \\
  amorphous carbon & ~12  & 1.85 & $2 \cdot 10^{4}$ & $3.3 \cdot 10^{-4}$ & 10 \\
  Mg$_2$SiO$_4$    & 140  & 3.27 & $2 \cdot 10^{1}$ & $3.6 \cdot 10^{-5}$ & $6 \cdot 10^{-3}$ \\
  MgFeSiO$_4$      & 172  & 3.71 & $7 \cdot 10^{3}$ & $3.6 \cdot 10^{-5}$ & 2 \\
  & \\
  \hline
  & \\
\end{tabular}

\noindent As expected, the value of $\Gamma$ for amorphous carbon
grains is well above the threshold for a dust-driven wind, even for
a degree of condensation $f_c \approx 0.3-0.5$ as typically found in
detailed models. For silicate grains, the picture is, again, more
complex: for forsterite (Mg$_2$SiO$_4$), on the one hand, $\Gamma$
is so far below the critical value that even the scenario of a
pulsation-supported dust-driven outflow as discussed above is
unrealistic, since it would require $\left( u_{\rm shock} / u_{\rm
esc} \right)_{\rm r=R_{\ast}}$ close to unity. Silicate grains
containing equal amounts of Mg and Fe, on the other hand, could
result in a sufficiently high $\Gamma$, but as we saw above they
will not form sufficiently close to the star to drive a wind.

At this point one might wonder about alternatives to the ballistic
one-shock scenario for the trajectory prior to dust condensation. It
might be possible that a matter element gets hit by several shocks
at increasing distance from the star if the fall-back is slow enough
('parachute regime', see above), gradually driving the gas out to a
considerable distance (see, e.g., Bowen 1988). One could even think
about replacing the first step(s) on the way with other mechanisms
which on their own may not be sufficient to drive an outflow (just
like the shock waves alone will hardly do the trick, at least not
within observed constraints as known today). Several such
possibilities have been discussed in the literature earlier, e.g.,
Alfv{\'e}n waves (e.g., Hartmann \& MacGregor 1980), sound waves
(e.g., Pijpers \& Hearn 1989), or high thermal gas pressure due to
heating by shock waves ('calorisphere', Willson \& Bowen 1998).
However, each of these alternatives faces the same question: will
densities and dynamical timescales allow for efficient grain
condensation at a certain distance from the star?

\subsection{Condensation efficiency and time scales}

Discussions about dust in atmospheres and winds of AGB stars often
ignore the fact that a low enough temperature is only a necessary
condition for the formation or survival of dust grains, and not a
sufficient one. In dynamical atmospheres and winds where timescales
are set by pulsation and wind dynamics, the efficiency of dust
formation is strongly dependent on prevailing gas densities and
abundances. As matter moves away from the star, the temperature
decreases, which -- at first -- favors grain formation. At the same
time, however, the growth of grains turns into a race against
falling density which slows down the process, and more often than
not leads to incomplete condensation.

Simple estimates based on gas kinetics as presented in Gustafsson \&
H{\"o}fner (2004) demonstrate that the timescales for the growth of
carbon grains close to the condensation radius (i.e. at about 2-3
stellar radii) are on the order of a year, i.e. comparable with the
pulsation period of the star, and increasing outwards with falling
density. This is in good agreement with detailed numerical models
which tend to show mean degrees of condensation well below unity
(typically 0.3-0.5), and a rather limited zone of grain growth.

Iron-poor silicates, which should have a condensation radius similar
to amorphous carbon grains (see above), face the problem that the
abundance of the key element Si is about an order of magnitude lower
than that of C, increasing the estimate for the grain growth time by
a corresponding factor (the timescale is inversely proportional to
this abundance). This might result in a rather low dust formation
efficiency, which, in combination with the low opacity of such
grains, makes them a by-product of the wind, not a driver.

Iron-rich silicates, on the other hand, which could in principle
contribute to the total opacity, and consequently to driving the
wind, will most likely not form in significant amounts, being
handicapped by both the low abundance of Si and the much lower
densities prevailing at distances corresponding to their large
condensation radius. At such distances, the densities will be at
least an order of magnitude lower than in the region where carbon or
iron-poor silicates may form, adding another factor of ten to the
grain growth timescale.

\section{Detailed models: status and trends}

The development of models for a particular phenomenon -- in this
case mass loss through dusty winds -- often occurs in several steps:
first ideas about basic processes lead to order of magnitude
estimates, followed by simple (analytical or numerical) models. If
the basic principles seem sound the next step is an iterative
improvement of numerical models by comparison with observations,
leading eventually to detailed, reasonably realistic models which,
finally, can be applied to study certain astrophysical phenomena in
a wider context (e.g. the role of dusty winds in stellar and
galactic evolution, to pick a not-so-random example).

Detailed time-dependent models for winds of AGB stars currently come
in three major groups, namely two types of spherically symmetric
frequency-dependent models including non-equilibrium dust formation
for C-rich and O-rich chemistry, respectively, and 2D/3D models
concerned with the effects of giant convection cells and structure
formation on atmospheres and mass loss. These three types of models
have reached different stages in their development, as will be
discussed below.\footnote{Other kinds of models, such as stationary
wind models, neglecting pulsation and shocks, or pure dynamical
atmosphere models without mass loss (e.g. Bessel et al. 1996, Scholz
\& Wood 2000, Tej et al. 2003ab), are not included in this
discussion. Note also that the recent models by Ferrarotti \& Gail
(2006) studying dust formation for a wide range of stellar
parameters and chemical compositions are not wind models in the
strict sense of the word since they have mass loss rates as an input
parameter, not as a result.}

Early pioneering models in this field, exploring the general effects
of pulsations and dust for cool stellar winds (e.g. Wood 1979, Bowen
1988) make hardly a distinction between C-rich and O-rich stars,
with the possible exception of choosing appropriate values for
certain input parameters. The turning point came with the inclusion
of a detailed description of dust formation (e.g., Fleischer et
al.~1992, H{\"o}fner \& Dorfi 1997, Winters et al. 2000, Jeong et
al. 2003) in contrast to a parameterized description of the dust
opacity, and/or the introduction of frequency-dependent radiative
transfer, accounting for the complex opacities of molecules and dust
(e.g., H{\"o}fner 1999, Woitke 2006a).

The latest generation of models for atmospheres and winds of C-rich
AGB stars by H{\"o}fner et al.~(2003), combining a
frequency-dependent treatment of radiative transfer with
time-dependent hydrodynamics and a detailed description of dust
formation, compares well with various types of observations, such as
low-resolution NIR spectra (Gautschy-Loidl et al. 2004) or profiles
of CO vibration-rotation lines (Nowotny et al. 2005ab and this
volume). With these non-grey dynamical models it is possible for the
first time to simultaneously reproduce the time-dependent behavior
of fundamental, first and second overtone vibration-rotation lines
of CO, features originating in the outflow, dust formation region,
and pulsating atmosphere, respectively, probing the dynamics from
the photosphere out into the wind. Recently, Mattsson et al. (2007)
have applied these models to investigate the formation of detached
shells in connection with a He-shell flash. Currently, a large grid
of dust-driven wind models for C-rich AGB stars and an accompanying
library of variable synthetic spectra are being computed (see
Mattsson et al., this volume).

The development of similar detailed models for winds of O-rich AGB
stars has been lagging behind the C-rich case, not the least due to
a more complex scenario for dust formation. Jeong et al. (2003)
presented wind models for M-type stars, combining a detailed
description of dust formation with time-dependent dynamics and grey
radiative transfer. While the stellar parameters of these models are
somewhat on the extreme side (high luminosities, low effective
temperatures), the resulting wind characteristics are reasonably
realistic. In view of the discussion in the previous section, it may
be surprising that it is possible to drive winds with silicate
grains. This is a direct consequence of the grey radiative transfer
used in these models which corresponds to an effective value of
$p=0$ (cf. Fig.~\ref{f_rcond}), resulting in a rather small
condensation radius for all types of grains, including iron-rich
species. Frequency-dependent models for O-rich AGB stars by Woitke
(2006a, and this volume) clearly demonstrate that the iron content
of silicate grains has to be very low (large condensation radius for
iron-rich silicates, combined with low condensation efficiency, see
discussion in previous section), and that the wind -- even for quite
extreme stellar parameters -- will consequently not be driven by
silicate grains, in contrast to previous expectations.

During recent years, 2D/3D dynamical atmosphere and wind models for
AGB stars have emerged, in addition to the spherically symmetric
models discussed above. The computational effort behind such models
is considerable, and several simplifications have to be introduced
in the description of physical processes. Nevertheless,
investigating the effects of intrinsically three-dimensional
phenomena like convection or flow instabilities on mass loss, seems
a timely project in view of recent interferometric observations
which indicate deviations from spherical symmetry. Woitke (2006b)
presented 2D (axisymmetric) dust-driven wind models, including
time-dependent dust formation and grey radiative transfer. He
studied how instabilities in the dust formation process create
intricate patterns in the circumstellar envelope, but without taking
the pulsation of the central star into consideration. Freytag \&
H{\"o}fner (2003, 2007), on the other hand, investigate the effects
of giant convection cells and of the resulting shock waves in the
atmosphere on time-dependent dust formation in the framework of 3D
RHD 'star-in-a-box' models. The atmospheric patterns created by
convective motions are found to be reflected in the circumstellar
dust distribution, due to the strong sensitivity of grain formation
to temperatures and gas densities, as discussed above.

\section{Conclusions}

The well-known dichotomy between M-type and C-type AGB stars, as
observed in molecular spectra, may have an even more drastic
consequence for their mass loss mechanism. While advances in
modelling, in particular the introduction of frequency-dependent
radiative transfer in time-dependent dynamical models, have improved
agreement between models and observations of C-rich AGB stars, the
opposite seems to be true for the O-rich case. Recent models of
M-type stars combining time-dependent dust formation with
frequency-dependent radiative transfer demonstrate that silicate
grains forming in such environments will be extremely iron-poor,
resulting in too low opacities to drive a wind.

These qualitative differences can be understood with simplified
analytical considerations, as discussed here, causing serious doubt
about the validity of the dust-driven wind scenario for M-type AGB
stars, at least in its most simple form. The mystery is deepened by
the fact that the observed wind characteristics for both types of
stars are rather similar (see, e.g., Olofsson 2004, Ramstedt et al.
2006), which hints at a common mass loss mechanism. In addition,
alternative scenarios discussed in the literature may have serious
difficulties explaining the formation of considerable amounts of
dust as a by-product in an outflow driven by a different force.
Non-equilibrium dust condensation is very sensitive to the
prevailing thermodynamical conditions, and restricted to a
relatively narrow zone close to the star, putting strong constraints
on potential driving forces.

In this situation the role of observers should not be
underestimated. Any observations which can narrow down the possible
range of conditions in the wind acceleration zone are of great
importance for solving this problem.

\end{document}